\documentclass{SciPost}
\usepackage{subfig}
\hypersetup{
    colorlinks,
    linkcolor={red!50!black},
    citecolor={blue!50!black},
    urlcolor={blue!80!black}
}

\usepackage[bitstream-charter]{mathdesign}
\DeclareSymbolFont{usualmathcal}{OMS}{cmsy}{m}{n}
\DeclareSymbolFontAlphabet{\mathcal}{usualmathcal}

\fancypagestyle{SPstyle}{
\fancyhf{}
\lhead{\colorbox{scipostblue}{\bf \color{white} ~SciPost Physics Proceedings }}
\rhead{{\bf \color{scipostdeepblue} ~Submission }}

\fancyfoot[C]{\textbf{\thepage}}
}

\begin{document}

\pagestyle{SPstyle}

\begin{center}{\Large \textbf{\color{scipostdeepblue}{
The magic of top quarks\\
}}}\end{center}

\begin{center}\textbf{
Chris D. White\textsuperscript{1$\star$} and
Martin J. White\textsuperscript{2$\dagger$}
}\end{center}

\begin{center}
{\bf 1} Queen Mary University of London, London, UK
\\
{\bf 2} University of Adelaide, Adelaide, Australia
\\[\baselineskip]
$\star$ \href{mailto:email1}{\small christopher.white@qmul.ac.uk}\,,\quad
$\dagger$ \href{mailto:email2}{\small martin.white@adelaide.edu.au}
\end{center}

\definecolor{palegray}{gray}{0.95}
\begin{center}
\colorbox{palegray}{
  \begin{tabular}{rr}
  \begin{minipage}{0.36\textwidth}
    \includegraphics[width=60mm,height=1.5cm]{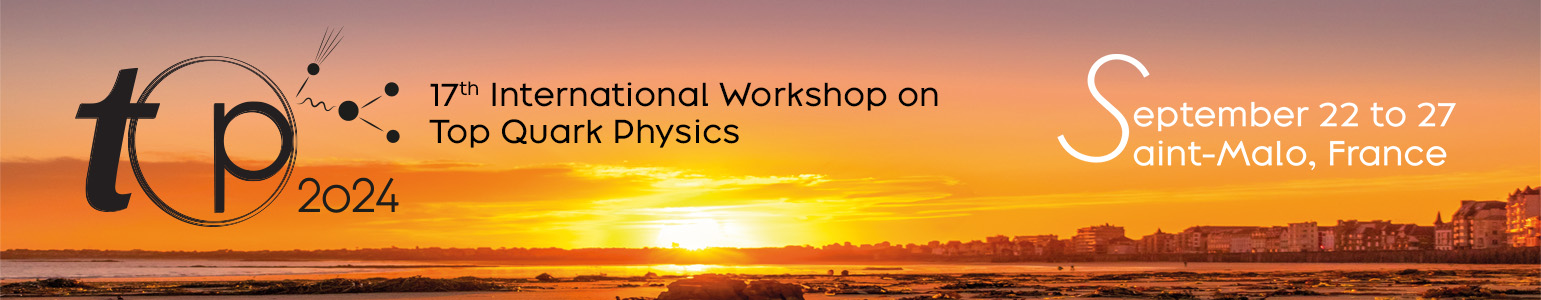}
  \end{minipage}
  &
  \begin{minipage}{0.55\textwidth}
    \begin{center} \hspace{5pt}
    {\it The 17th International Workshop on\\ Top Quark Physics (TOP2024)} \\
    {\it Saint-Malo, France, 22-27 September 2024
    }
    \doi{10.21468/SciPostPhysProc.?}\\
    \end{center}
  \end{minipage}
\end{tabular}
}
\end{center}

\section*{\color{scipostdeepblue}{Abstract}}
\textbf{\boldmath{%
In recent years, there has been increasing collaboration between the
fields of quantum computing and high energy physics, including using
LHC processes such as top (anti-)quark pair production to perform high
energy tests of quantum entanglement. In this proceeding, I will review
another interesting property from quantum computing (“magic”), that is
needed to make quantum computers with genuine computational advantage
over their classical counterparts. How to make and enhance magic in
general quantum systems is an open question, such that new insights
are always useful. To this end, I will show that the LHC naturally
produces magic top quarks, providing a novel playground for further
study in this area.
}}

\vspace{\baselineskip}

\noindent\textcolor{white!90!black}{%
\fbox{\parbox{0.975\linewidth}{%
\textcolor{white!40!black}{\begin{tabular}{lr}%
  \begin{minipage}{0.6\textwidth}%
    {\small Copyright attribution to authors. \newline
    This work is a submission to SciPost Phys. Proc. \newline
    License information to appear upon publication. \newline
    Publication information to appear upon publication.}
  \end{minipage} & \begin{minipage}{0.4\textwidth}
    {\small Received Date \newline Accepted Date \newline Published Date}%
  \end{minipage}
\end{tabular}}
}}
}


\vspace{10pt}
\noindent\rule{\textwidth}{1pt}
\tableofcontents
\noindent\rule{\textwidth}{1pt}
\vspace{10pt}


\section{Introduction}
\label{sec:intro}

In recent years, an increasing number of people have looked at using
high energy colliders to test fundamental properties of quantum
mechanics. Perhaps the most well-known quantum concept is that of {\it
  entanglement}, which clearly distinguishes quantum behaviour from
classical, as encoded e.g. via Bell inequalities~\cite{Bell:1964kc}. A
particularly useful system for studying entanglement at the Large
Hadron Collider is that of top (anti-)top pair production, examined in
this context in
e.g. refs.~\cite{Afik:2020onf,Afik:2022kwm,Dong:2023xiw,Fabbrichesi:2021npl,Aoude:2022imd,Aguilar-Saavedra:2022uye}
(see also ref.~\cite{Barr:2024djo}, and ref.~\cite{Abel:1992kz} for a
critical appraisal of such measurements). Entanglement is, however,
not the only special property of quantum states. Lots of other things
are studied in either Quantum Computation or Information theory, for a
variety of interesting reasons. Might these also be useful for high
energy physics? In this proceeding, we examine one such property --
{\it magic} -- and argue that the LHC indeed offers an interesting
situation for studying it. To introduce magic, let us first briefly
review aspects of quantum computing.

\section{A bit of quantum computing}
\label{sec:QC}

In quantum computers, classical bits (with values $\{0,1\}$) are
replaced with {\it qubits}, namely normalised quantum states
\begin{equation}
  |\psi\rangle=\alpha|0\rangle+\beta|1\rangle,\quad
  |\alpha|^2+|\beta|^2=1,
  \label{qubit}
\end{equation}
where $|0\rangle$ and $|1\rangle$ are orthogonal basis states, and
$\alpha$, $\beta$ complex coefficients. The canonical example of a
single qubit system is a spin-1/2 particle, in which case $|0\rangle$
and $|1\rangle$ may represent the two linearly independent spin
states. Multi-qubit systems can then be described using a basis
comprised of tensor products of single-qubit states. Quantum
computers take (multi-)qubits, and subject them to unitary
transformations, where unitarity corresponds to conservation of
probability in quantum theory. Each transformation acts on a given
number of qubits, and is known as a {\it quantum gate}. These are the
equivalent of {\it logic gates} in classical computing, and the
quantum versions have fancy names such as {\it Hadamard}, {\it phase},
{\it CNOT} and {\it Pauli}. Precise details may be found e.g. in
ref.~\cite{Nielsen:2012yss}, or ref.~\cite{White:2024nuc} in the
context of this proceeding.

\section{Could it be magic?}
\label{sec:magic}
Quantum computers are expected to vastly outperform their classical
counterparts, which is na\"{i}vely due to the two quantum properties
of {\it superposition} and {\it entanglement}. However, it turns out
that this is not quite true, and to see why, we need the concept of a
{\it stabiliser state}. These are multiqubit states that give a simple
spectrum for a restricted set of operators known as {\it Pauli
  strings}:
\begin{equation}
  {\cal P}_n=P_1\otimes P_2\otimes\ldots\otimes P_N,\quad
  P_a\in\{\sigma_1^{(a)},\sigma_2^{(a)},\sigma_3^{(a)},I^{(a)}\}.
  \label{Pndef}
\end{equation}
In words: a Pauli string acting on $n$ qubits operates on the $a^{\rm
  th}$ qubit with a Pauli matrix, or an identity matrix. There are
then $4^n$ such strings for $n$ qubits. Stabiliser states are such
that the set of expectation values of all Pauli strings have $2^n$
values which are $\pm 1$, and the rest zero. This is in contrast to a
general state, which will have a wide variation of expectation values
of each Pauli string. We can make these stabiliser states by acting on
the state $|0\rangle\otimes|0\rangle\otimes\ldots\otimes|0\rangle$
with the particular set of quantum gates listed above.

To the uninitiated, the above definition will be utterly opaque. But
it becomes important due to something known as the {\it
  Gottesman-Knill theorem}~\cite{Gottesman:1998hu}. Roughly speaking,
this states that for any quantum computer containing stabiliser states
only, there is a classical computer that is just as efficient!
Stabiliser states can include certain maximally entangled
states. Thus, something other than entanglement is needed for
efficient quantum computing.

The ``something else'' has been called {\it magic} in the literature,
and from the above comments basically measures ``non-stabiliserness''
of a quantum state. Different definitions of magic exist in the
literature, and we will here use the {\it Stabiliser R\'{e}nyi
  Entropies} of ref.~\cite{Leone:2021rzd}:
\begin{equation}
  M_q=\frac{1}{1-q}\log_2(\zeta_q),\quad \zeta_q\equiv
  \sum_{P\in{\cal P}_n} \frac{\langle\psi|P|\psi\rangle^{2q}}{2^n}.
    \label{Mqdef}
\end{equation}
In these formulae, $q\geq 2$ is an integer, and $\zeta_q$ represents a
weighted sum over the Pauli spectrum values raised to a power. We can
think of the set of values $\{M_q\}$ as providing moments of the Pauli
spectrum, and this definition turns out to have the following
desirable properties: (i) the magic is additive when combining quantum
systems; (ii) it vanishes for stabiliser states. In what follows, we
will focus on $q=2$ (i.e. the {\it Second Stabiliser R\'{e}nyi
  Entropy}), given this is already sufficient to quantify non-zero
magic. We now have everything we need to investigate magic at the LHC!

\section{Are top quarks magic?}
\label{sec:tops}

\begin{figure}
    \centering
    \subfloat[]{\includegraphics[width=0.45\textwidth]{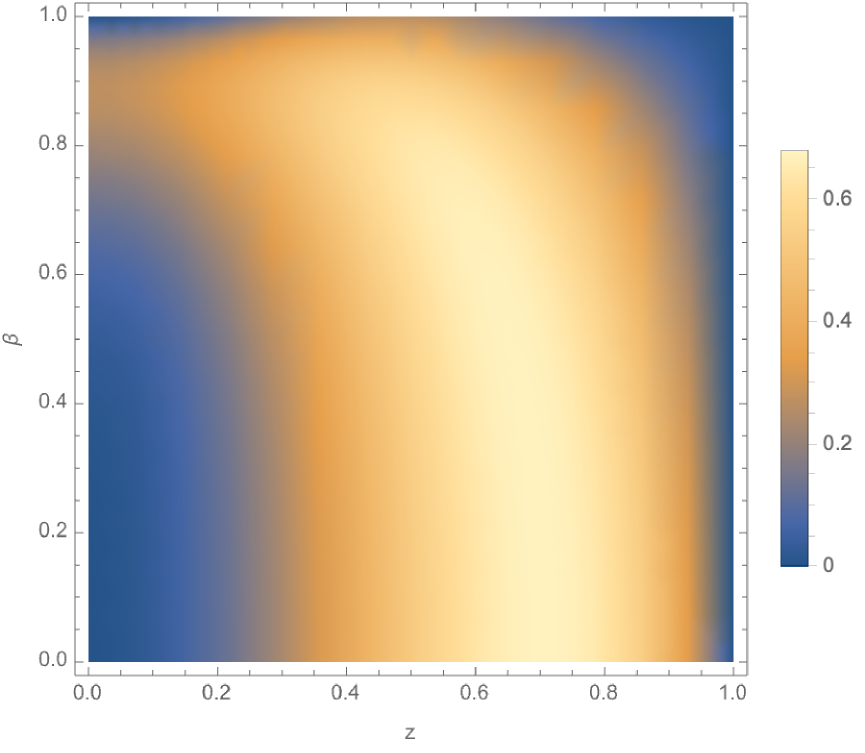} }
    \qquad
    \subfloat[]{\includegraphics[width=0.45\textwidth]{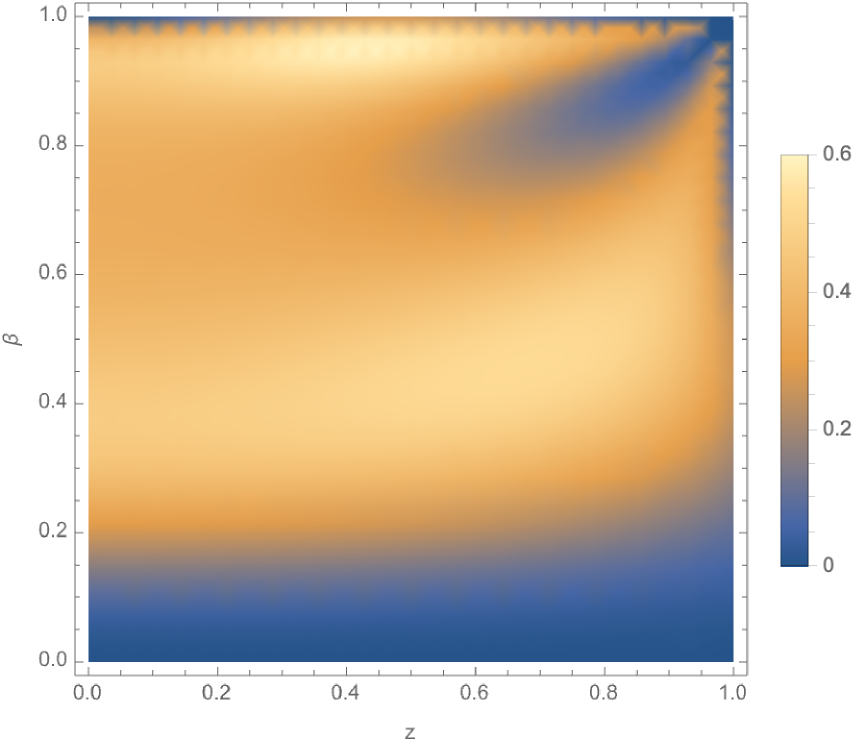} }
    \caption{The magic of a mixed top-antitop final state in: (a) the
      $q\bar{q}$ channel; (b) the $gg$ channel.}
    \label{fig:magic}
\end{figure}

Given its previous success in probing entanglement, we can look at top
pair production as a potential playground for exploring magic. The
Standard Model tells us that the most general configuration of
top-antitop spins in pair production at the LHC is a {\it mixed state}
(i.e. a superposition of so-called {\it pure states}). Such states can
be described using the density matrix formalism, and the spin density
matrix for a top-antitop pair in partonic channel $I$ has the general
form
\begin{equation}
  \rho^I\sim \tilde{A}^I I_4+\sum_i\left(
  \tilde{B}_i^{I+}\sigma_i\otimes I_2+\tilde{B}^{I-}I_2\otimes \sigma_i\right)
  +\sum_{i,j}\tilde{C}_{ij}\sigma_i\otimes \sigma_j,
  \label{rhoIdef}
\end{equation}
where $I_n$ is an $n$-dimensional identity matrix. The quantities
$\{\tilde{A}^I,\tilde{B}^{I\pm}, \tilde{C}_{ij}\}$ are called {\it
  Fano coefficients}, and are respectively related to the total
cross-section, polarisation of the (anti-)top, and spin
correlations. Each coefficient depends upon the invariant mass and
scattering angle of the top particles, as well as the basis chosen to
relate the spin directions $(1,2,3)$ to three directions in physical
space. A common choice is the {\it helicity
  basis}~\cite{Baumgart:2012ay}, in which one chooses an orthogonal
coordinate system aligned with the top quark direction. Defining the
normalised Fano coefficients via
\begin{equation}
  B_i^{I\pm}=\frac{\tilde{B}^{I\pm}}{\tilde{A}^I},\quad
    C_{ij}^{I}=\frac{\tilde{C}^{I}_{ij}}{\tilde{A}^I},
      \label{BCnorm}
\end{equation}
the magic of a top quark pair is given by
\begin{equation}
  \tilde{M}_2(\rho^I)=
  -\log_2\left(\frac{1+\sum_{i}[(B_i^{I+})^4+(B_i^{I-})^4]
    +\sum_{i,j}(C_{ij}^I)^4}
  {1+\sum_{i}[(B_i^{I+})^2+(B_i^{I-})^2]
    +\sum_{i,j}(C_{ij}^I)^2}
  \right).
  \label{M2tilderes3}
\end{equation}
Figure~\ref{fig:magic} shows the magic, as calculated in the SM, for
both the $q\bar{q}$ and $gg$ initial states. We see that the magic is
concentrated away from extreme kinematic limits (e.g. threshold / high
energy), which is not surprising: it is known that the top quark final
state becomes separable and / or maximally entangled in these
regions. These happen to be stabiliser states, and thus the magic
vanishes. Note also that the magic can be non-zero where entanglement
vanishes, which does not contradict the fact that both entanglement
and magic are needed for quantum computational advantage: the latter
is a statement about {\it algorithms} or {\it circuits}, which must
necessarily contain both entangled and magic states in some
intermediate step(s). A given intermediate state, however, does not
need to be both entangled and magic.

Results for the magic are also shown for proton-proton initial states
in ref.~\cite{White:2024nuc}. Combining partonic channels or averaging
over angles typically increases magic, due to having a more mixed
state, whose Pauli spectrum becomes more complex as a result.

\section{Conclusion}
\label{sec:conclusion}

How to produce and enhance magic in arbitrary quantum systems remains
an open research question. We have shown that top quarks provide a
system in which magic can be produced, and highly effectively studied
using event selection. This may provide insights into how to make
other magic systems. Optimistically, one might hope that magic could
be used in probing new physics, or in strengthening the already active
dialogue between Quantum Computing / collider physics. Further work to
investigate these questions is ongoing.

\section*{Acknowledgements}
This work was carried out with Martin White, and we are grateful to
Rafael Aoude and Hannah Banks for collaboration on related topics.


\paragraph{Funding information}
This work was supported by the UK Science and Technology Facilities
Council (STFC) Consolidated Grant ST/P000754/1 ``String theory, gauge
theory and duality'', and the Australian Research Council grants
CE200100008 and DP220100007.

\bibliography{refs.bib}


\end{document}